\documentclass[a4paper]{article}

\usepackage{icrc2013}
\usepackage[english]{babel}

\title{Recent VERITAS Results on VHE Gamma-ray Sources in Cygnus}

\shorttitle{VERITAS Sources in Cygnus}

\authors{
Rene A. Ong$^{1}$,
for the VERITAS Collaboration.
}

\afiliations{
$^1$ Department of Physics and Astronomy, University of California, 
Los Angeles, CA 90095, USA \\
\scriptsize{
}
}

\email{rene@astro.ucla.edu}

\abstract{
The Cygnus region of the Galactic plane is a promising target for high-energy and 
very high energy (VHE) gamma-ray telescopes as it is home to many potential sources, such 
as supernova remnants, pulsar wind nebulae, X-ray binaries and massive star clusters.
The VHE gamma-ray observatory VERITAS (Very Energetic Radiation Imaging Telescope 
Array System) is an array of four 12 m diameter imaging atmospheric Cherenkov telescopes 
located at Mt. Hopkins, AZ, USA. Over the period of 2007 to 2012, VERITAS has carried out 
extensive observations in the direction of Cygnus.  These observations were initiated by 
a sky survey that covered Galactic longitudes between 67 and 82 degrees and Galactic 
latitudes between -1 and 4 degrees. Additional deep observations have been made near
specific sources, including TeV J2032+4130, Cygnus X-3, VER J2019+407 
(SNR G78.2+2.1/$\gamma$-Cygni), 
CTB 87, and MGRO J2019+37. This paper summarizes the latest VERITAS results
on the various source detections in the direction of Cygnus and our current understanding
of the origin for the VHE gamma-ray emission.
}


\begin{document}
\maketitle

%Begin a section.
\section{Introduction}
Cosmic ray particles span an enormous energy range, from below 1 GeV to above $10^{20}$ eV, and
they pervade our Galaxy with an energy density comparable to that in the Galactic magnetic field or
starlight.  In spite of many years of research, the origin of the very high energy 
(VHE, E $>$ 100 GeV) cosmic rays remains a significant
mystery today. Neutral messengers, such as gamma rays and neutrinos can provide critical
information about cosmic ray origins as they directly trace the energetic processes taking place
at acceleration sites throughout our Galaxy.

In recent years, VHE gamma-ray telescopes have been very successful at detecting sources in our Galaxy
that can help to shed light on the origin of cosmic rays.
At the present time, there are $\sim$100 firmly established objects within a few degrees of the
Galactic plane \cite{bib:TeVCAT}.
Identified sources include:
1) supernova remnants (SNRs), that are generally believed to produce the dominant portion of the
cosmic rays up to the knee in the spectrum (i.e. up to E $\sim$1 PeV),
2) binary systems, in which the acceleration may involve colliding winds or accretion-powered jets,
3) pulsar wind nebulae, in which the acceleration results from the interaction of a pulsar-driven wind
with ambient radiation fields, and 
4) star forming regions, in which the gamma rays trace interactions taking place in molecular
clouds or near OB associations.
In addition to these sources, there are a significant number of VHE detections that are not
yet firmly associated with any known astronomical object. Understanding the nature 
of the unidentified VHE detections is thus an important goal in piecing together the
overall high-energy Galactic source population.

The Cygnus region of the Galactic plane is a prime target for high-energy and VHE gamma-ray
telescopes.  It is a region that is rich in star formation activity and it
contains a substantial catalog of possible accelerators of VHE cosmic rays.
The direction towards Cygnus contains three arms of the Milky Way galaxy in our line of sight,
making it difficult to determine counterparts to VHE gamma-ray sources in some cases, but leading
to a fertile region for exploration.
The region has been surveyed at GeV energies by EGRET and Fermi-LAT, at
TeV energies by HEGRA \cite{bib:HEGRA1}, and 
at multi-TeV energies by the Milagro air shower array \cite{bib:Milagro1,bib:Milagro2}.
Cygnus was also targeted by the VERITAS telescope which carried out a survey of the
region between 2007 and 2009.
The VERITAS Sky Survey is discussed in more detail below.

\section{VERITAS}
VERITAS (the Very Energetic Radiation Imaging Telescope Array System)
is a state-of-the-art ground-based VHE gamma-ray telescope. 
Located at the F. L. Whipple Observatory in southern Arizona, USA, VERITAS
uses the imaging atmospheric Cherenkov technique to detect gamma rays at
energies from 85 GeV to 30 TeV. The observatory consists of
an array of four large telescopes, each comprising a 
12 m diameter optical reflector and an associated camera spanning 
a field-of-view of 3.5$^\circ$. 
%Each telescope produces an image of the Cherenkov light generated
%in the atmosphere. A stereo reconstruction technique makes use of the 
%orientation, intensity and shape of each telescope image to estimate the direction,
%energy, and type of the primary particle (i.e. gamma ray or hadron).

Standard observations with the full four-telescope VERITAS array started
in September 2007 and the instrument has been upgraded successively since
then. In 2009, one of the telescopes was moved to give a more symmetrical
array geometry. The new geometry, combined with a reduced optical point
spread function, led to a significant improvement in the point-source 
sensitivity of VERITAS. The VERITAS upgrade, carried out between 2009 and
2012, led to a lowering of the energy threshold and a further improvement in
the sensitivity. 

\section{VERITAS Sky Survey}
The Cygnus region was targeted by the VERITAS Sky Survey, a flagship
project that was carried out by VERITAS between 2007 and 2009.
The base survey covered a region of $15^\circ \times 5^\circ$ in Galactic 
longitude and latitude, respectively. 
The survey encompassed ∼150 hours of observations at a uniform 
point-source sensitivity (99\% CL) of $< 4$\% 
of the Crab nebula flux. This sensitivity is a factor of five better 
than previously done \cite{bib:HEGRA1}.

Follow-up observations were carried out in subsequent seasons. The
combination of the base survey and the follow-up data led to the detection of
a number of sources plus interesting hints of sources (”hotspots”), 
including: 
1) an extended source near the shell-type SNR G78.2+2.1 ($\gamma$-Cygni), 
2) an extended source coinciding with the unidentified VHE gamma-ray source TeV
J2032+4130, and 
3) extended, and complex, emission in the region of the
unidentified Milagro source MGRO J2019+37.
The first object represented a new VHE source, 
named VER J2019+407, that is associated with the $\gamma$-Cygni SNR. 
The second and third objects were known TeV gamma-ray emitters, 
originally detected by HEGRA \cite{bib:HEGRA2} and Milagro,
respectively. 
For these two objects, the VERITAS observations 
represent the most sensitive exposures taken at these energies, 
and they have led to sharper views and better understanding of the sources.
Preliminary results from the VERITAS Sky Survey have been reported 
previously \cite{bib:Weinstein,bib:Ong}.
% and comprehensive results are the
%subject of a future publication.
This paper summarizes the updated results from these three source regions.

\section{VER J2019+407}
After the evidence from the VERITAS Sky Survey for a possible new source 
near the SNR G78.2+2.1 ($\gamma$-Cygni), VERITAS took follow-up data in the region in 2009.
Based on 21.4 hr of new data, a clear detection of a source was made at
a post-trials significance of 7.5 standard deviations ($\sigma$) \cite{bib:Aliu1}.
As shown in Figure~1, the VERITAS source (named VER J2019+407) 
has moderate extension and is displaced from the location of the gamma-ray pulsar
PSR J2021+4026.  The source does lie, however, on the northwest rim of the SNR, in
a region of enhanced radio emission.  GeV gamma-ray emission from the region has
been detected by Fermi-LAT; the center of this emission is displaced from the
VERITAS source, but its large extent fully encompasses the VHE gamma-ray source.
A maximum-likelihood fit to the VERITAS binned counts map yields a fitted
extension of $0.23 \pm 0.03 (stat) \, ^{+0.04} _{-0.02} (sys) $ degrees and
a fitted centroid position for the VHE source of
R.A. 20h 20m 04.8s, decl. +40$^\circ$ 45' 36" (J2000).

\begin{figure}[t]
  \centering
  \includegraphics[width=0.45\textwidth]{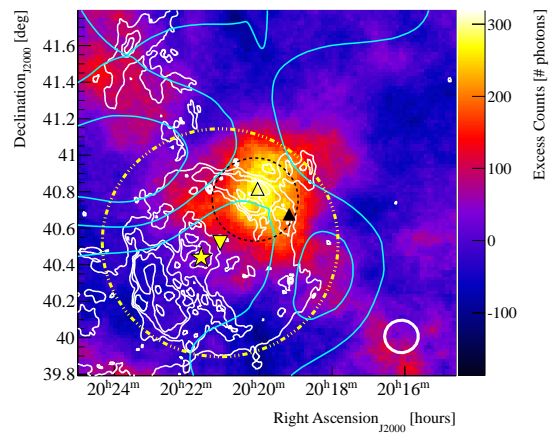}
  \caption{
VHE gamma-ray image of VER J2019+407 (color, indicating excess counts)
at energies above 320 GeV, overlaid with the 1420 MHz radio contours of the
SNR G78.2+2.1.
The fitted extent of the VHE source is shown by the black dashed
circle.  The yellow star gives the position of the pulsar PSR J2021+4026
and the inverted triangle and dashed yellow circle show the fitted centroid
and extent of the emission detected by Fermi-LAT above 10 GeV.
See \cite{bib:Aliu1} for details and references.
          }
\end{figure}

The differential gamma-ray energy spectrum for VER J2019+407 is determined
from a fit to those events within 0.24$^\circ$ of the source centroid.
Between 320 GeV and 10 TeV, the spectrum is well fit
by a single power law in energy with a differential spectral index
of $2.37 \pm 0.14_{stat} \pm 0.20_{sys}$ and a flux normalization at 1 TeV of
$1.5 \pm 0.2_{stat} \pm 0.4_{sys} \times 10^{-12}$ photons\,TeV$^{-1}$\,cm$^{-2}$\,s$^{-1}$.
The integral flux above 320 GeV corresponds to $\sim 3.7$\% of the Crab nebula
flux above that energy.
Analysis of data from both ROSAT and ASCA show that the VHE gamma-ray
source is coincident with enhanced X-ray emission with
the X-ray spectrum fit by a 
Raymond-Smith thermal plasma model \cite{bib:Aliu1}.
No evidence was found for significant non-thermal X-ray emission from the region near
VER J2019+407.

A variety of scenarios are possible when trying to explain the origin of the VHE
gamma-ray emission in the $\gamma$-Cygni region.
Many pulsar wind nebulae (PWNe) have been detected at very high energies, but
in the case of VER J2019+407, the VHE emission is significantly offset from
the pulsar PSR J2021+4026; the pulsar is not contained within the 99\% confidence
contour of the VHE source.
Conversely,
from the morphology of the radio, X-ray and VHE gamma-ray emission, it is more plausible
that the VHE emission arises from shock interactions between the 
ejecta of SNR G78.2+2.1 and the surrounding medium (in particular, a dense H I shell 
that appears to surround the SNR). For shocks that accelerate electrons to relativistic energies, 
the VHE gamma rays can result from inverse-Compton scattering of the electrons off 
ambient radiation fields.  Non-thermal X-ray emission is expected from synchrotron 
radiation of the same electrons, but the limit on this emission from our analysis 
is not low enough to constrain the possibility of
an inverse-Compton origin for the VHE gamma rays.

The VHE gamma-ray emission can also arise from the shock acceleration of hadrons
(particularly protons) that interact with target material to produce neutral pions
that then decay to gamma rays. A straightforward
calculation in the hadronic scenario determines
that an average target density of 1.0-5.5 cm$^{-3}$ is required to produce
the observed VHE gamma-ray flux; this density is
consistent with what we know about the region from other wavebands \cite{bib:Aliu1}.
In this calculation, estimates are made for the age and distance of the supernova
and a correction is made for the fraction of the remnant shell that appears to be
producing VHE gamma rays.

\section{TeV J2032+4130}
As the first unidentified source discovered at VHE gamma-ray energies,
as well as the first confirmed extended source,
TeV J2032+4130 holds a unique place in the field.
Detected first by HEGRA \cite{bib:HEGRA2}, it has also been studied at TeV
energies by Whipple \cite{bib:Whipple1} and 
MAGIC \cite{bib:MAGIC1}.  Within errors, the measurements made by the
atmospheric Cherenkov telescopes agree on the position of the source and
the HEGRA \cite{bib:HEGRA3} and MAGIC data indicate a source extension of $\sim$6 arc-minutes.
The source was also observed by the Milagro \cite{bib:Milagro1} and
ARGO-YBJ \cite{bib:ARGO1} air shower arrays, with Milagro detecting
bright emission from the region that is much larger in extent than seen for
TeV J2032+4130. The Milagro source was given the name MGRO J2031+41.
In addition to these observations, there has been extensive multiwavelength 
work to identify the counterpart of the source, but, in spite of all
of these efforts, no clear picture of its nature has emerged.

As discussed earlier, VERITAS saw evidence for TeV J2032+4130 from
data taken in the VERITAS Sky Survey between 2007 and 2009.
Here we discuss new results on the source from a deep exposure of the region
using data taken between 2009 and 2012 \cite{bib:Aliu2}.
This exposure, totalling 48.2 hr, represents the most sensitive study of TeV 2032+4130 yet
done at these energies.
With these data, the source is clearly detected at a significance
level of 8.7$\sigma$.
A fit to the excess counts map yields a centroid position
of R.A. 20h 31m 39.8s, decl. +41$^\circ$ 33' 53" (J2000) and
an asymmetric extension with a major axis of
$0.16 \pm 0.02$ degrees, oriented to the north east, and a minor
axis of $0.066 \pm 0.009$ degrees.  
The VERITAS source is named VER J2032+415.
Its position is consistent
with previous measurements but VERITAS sees an extension that is asymmetric 
and is somewhat larger than earlier results.
The energy spectrum is well fit
by a single power law in energy with a differential spectral index
of $2.05 \pm 0.16_{stat} \pm 0.21_{sys}$ and a flux normalization at 1 TeV of
$9.3 \pm 1.6_{stat} \pm 2.2_{sys} \times 10^{-13}$ photons TeV$^{-1}$\,cm$^{-2}$\,s$^{-1}$.
The integral flux above 1 TeV corresponds to $\sim 4.3$\% of the Crab nebula
flux above that energy.
There is no evidence for flux variability for the data taken over this
three-year period.  A careful study was made of the VHE gamma-ray image in
three energy ranges, but no evidence was seen of an energy-dependent morphology.

A blind search of the Fermi-LAT data revealed a previously unknown gamma-ray
pulsar, PSR J2032+4127, having a pulse period of 142 ms and located approximately
0.2 degrees from the center of VER J2032+415 \cite{bib:Camilo}. 
To look for unpulsed GeV emission from the region, we carried out an analysis
of the four-year data set of Fermi-LAT.  Selecting only those photons in
an off-pulse window and contained within 0.5$^\circ$ of the pulsar position,
we used a binned likelihood analysis to search for nebular emission.
No such emission was detected, allowing us to place flux upper limits 
in three energy bands \cite{bib:Aliu2}.

\begin{figure}[t]
  \centering
  \includegraphics[width=0.43\textwidth]{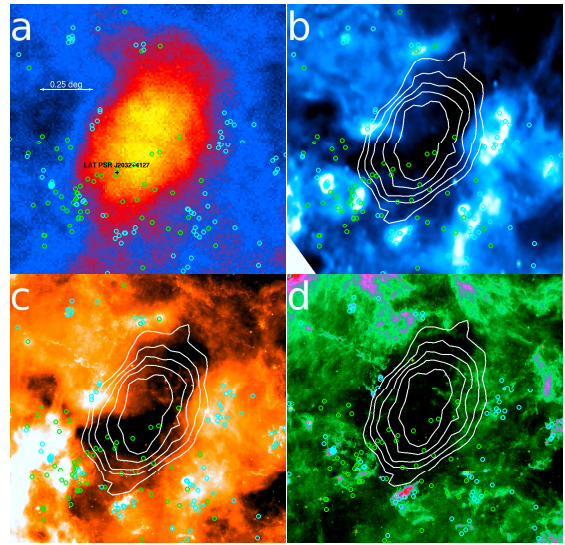}
  \caption{
Images of the region around VER J2032+415 in different wavelengths:
a) VHE gamma rays; VERITAS significance map showing position of the
Fermi-LAT pulsar and Cyg OB2 \#5,
b) 1.4 GHz from Canadian Galactic Plane Survey,
c) 24 $\mu$m from Spitzer MIPS,
and 
d) 8 $\mu$m from Spitzer GLIMPSE.  
In images b, c, and d, the contours indicate VERITAS significance from 4 to
8 standard devitations.  Green (cyan) circles indicate OB stars (star forming
regions). 
The angular size is shown in image a.
See \cite{bib:Aliu2} for details and references.
          }
\end{figure}

The Cygnus X complex, where VER J2032+415 is located, is one of the most
active star formation regions in the Galaxy.  In Figure~2, images of
the source and its surrounding region are shown in different wavebands.
The striking feature of this figure is that VER J2032+415 appears to be located
in a void in the generally bright diffuse emission at longer wavelengths.
The origin of this void is not clear; it could be due to a supernova explosion
and subsequent expansion as a supernova remnant or possibly due to stellar winds arising from the
Cygnus OB2 association.
A remnant shell comensurate with the size of the void has not been identified, but it might
be very faint for a relatively old SNR.
In this case, the VHE gamma-ray emission could arise from a pulsar wind 
powered by the Fermi-LAT pulsar PSR J2032+4127 and filling the interior of the SNR.
In this scenario, the corresponding X-ray and GeV gamma-ray pulsar nebulae must be too
faint to be detected.
Compared to other known TeV pulsar wind nebula (PWN) sources, 
the characteristics of VER J2032+415 are not
exceptional \cite{bib:Aliu2}.
Similarly, PSR J2032+4127
is one of the oldest and weakest pulsars seen in association with a VHE gamma-ray PWN, 
but is still within the population of pulsars from which TeV PWN are detected.

The VERITAS observations of the region are joined by those of the air shower arrays
Milagro \cite{bib:Milagro3} and ARGO \cite{bib:ARGO1}. 
Their observations resulted a significantly softer spectral index of 
$3.1 \pm 0.2$ and $2.8 \pm 0.4$, respectively, both with 
increased flux levels, while integrating over larger areas of the sky. 
ARGO operates at energies similar to those of the imaging Cherenkov 
telescopes and sees an extension which is roughly similar. However, MILAGRO begins its 
measurements at 20 TeV and observes a $3.0 \pm 0.9$ deg spatial extension. 
%Owing to the poorer angular resolution of MILAGRO, it integrates over a larger region of the
%sky than VERITAS and may include possible components coming from outside the source.
%However, when we integrate the VERITAS excess over a region corresponding to the Milagro point-spread-function,
%the flux seen by Milagro is still higher.
%When we integrate the VERITAS excess over a region of the MILAGRO psf, the flux seen by MILAGRO is still higher.
At GeV gamma-ray energies the $Fermi$-LAT has reported the existence of a 
$Cocoon$ located in the Cygnus region between the locations of  TeV J2032+4130 and the $\gamma$-Cygni SNR.  
In their paper \cite{bib:Cocoon}, they suggest that this excess is due to a population of freshly accelerated cosmic rays. 
Reconciling these differing sets of observations may well be the key to understanding this complex and interesting 
region. 

\section{MGRO J2019+37 Region}

In their survey of the northern hemisphere sky at multi-TeV energies, Milagro
discovered three relatively bright sources
(i.e. at a significant fraction of the Crab nebula flux):
MGRO J1908+06, MGRO J2031+41 (discussed earlier in the context of TeV J2032+4130), and
MGRO J2019+37 in the Cisne region \cite{bib:Milagro2}.
This third detection yielded a relatively broad ($1.2^\circ \times 0.7^\circ$) feature
that was not spatially resolved.
The excellent angular resolution of VERITAS enables a sharper view of this VHE bright
region of the sky.

Dedicated VERITAS observations of the region took place in Fall 2010, totalling
71 hr. 
These observations yield a complex sky map in which multiple VHE gamma-ray sources
are clearly evident \cite{bib:Aliu3}.
VERITAS detects broad and extended emission from a region that
is largely contained by the error circle of MGRO J2019+37 and separate and fainter 
emission that is consistent spatially with a point source.
These two sources are given the names VER J2019+368 and VER J2016+371 \cite{bib:Aliu4}.

The statistical significance of VER J2016+371 is 5.7$\sigma$
(post-trials) and the
fitted source centroid is positionally consistent with the SNR CTB 87.
The energy spectrum is well fit between 650 GeV and 20 TeV
by a single power law in energy with a differential spectral index
of $2.3 \pm 0.3_{stat} \pm 0.3_{sys}$ and a flux normalization at 1 TeV of
$3.1 \pm 0.9_{stat} \pm 0.6_{sys} \times 10^{-13}$ photons\,TeV$^{-1}$\,cm$^{-2}$\,s$^{-1}$.
A multiwavelength study \cite{bib:Aliu4} of CTB 87 indicates that it is most likely a PWN
that is seen in both X-rays and 
VHE gamma rays.
The properties (age, high-energy luminosity, etc.) of CTB 87 place it in line
with other PWN detected at TeV gamma-ray energies.

\begin{figure}[t]
  \centering
  \includegraphics[width=0.45\textwidth]{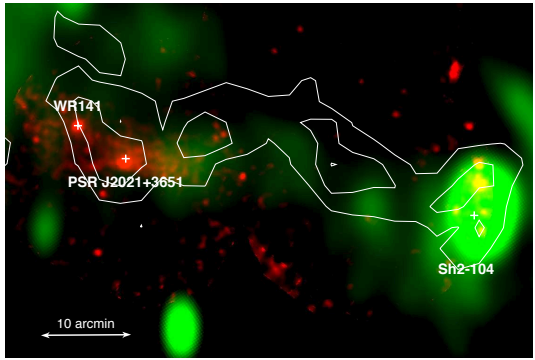}
  \caption{
The inner region of MGRO J2019+37 as seen in radio (green) and
X-ray (red).  The VERITAS emission associated with the
source VER J2019+368 is shown as the white contours with
significance levels of 5 and 4$\sigma$.
The positions of the pulsar PSR J2021+3651, the Wolf-Rayet star WR141,
and the HII region Sh2-104 are also indicated.
See \cite{bib:Aliu4} for more details.
          }
\end{figure}

The region encompassing VER J2019+368 is more complex and it is likely that this
object is made up of multiple sources that extend approximately 
one degree across the sky.
The VHE energy spectrum is estimated from a circular region of radius 0.5 degs centered
on the best fit position of VER J2019+368.
The resulting spectrum is well fit between 1 TeV and 30 TeV by a single
power law in energy with a hard differential spectral index of
$1.75 \pm 0.08_{stat} \pm 0.3_{sys}$ and a flux normalization at 1 TeV of
$1.4 \pm 0.1_{stat} \pm 0.3_{sys} \times 10^{-12}$ photons\,TeV$^{-1}$\,cm$^{-2}$\,s$^{-1}$.

Understanding the origin of the VHE emission 
in VER J2019+358 is challenging and there are a number of
objects in the region that may be connected to this emission, including
the energetic pulsar PSR J2021+3651, the Wolf-Rayet star WR141, a
transient Integral source IGR J20188+3647, and the HII region Sh2-104.
An extensive multiwavelength study of the region has been carried out \cite{bib:Aliu4}.
Although no unambiguous counterparts to the VHE gamma-ray emission have been identified,
the emission tends to follow a ridge of diffuse radio emission going east from SH2-104
to PSR J2021+3651, as shown in Figure 3.  It is plausible that a substantial fraction
of the VHE emission derives from a PWN powered by PSR J2021+3651, leaving the remaining
fraction as yet unexplained.

\section{Conclusions}

Because of its high star-forming activity and concentration of potential sources,
the Cygnus region is perhaps the best location in the northern hemisphere to study
processes leading to VHE gamma-ray emission with the goal of understanding the
origin of Galactic cosmic rays.
VERITAS has made extensive observations in the region, starting with a sky survey
between 2007 and 2009 and continuing with dedicated follow-up observations in
the following years.
The observations have so far yielded a number of significant results,
including the discovery of a new source VER J2019+407,
more precise measurements of the first unidentified source
TeV J2032+4130, and a significantly
sharper view of the complex region near MGRO J2019+37.
Future work will focus on a comparison of the VHE gamma-ray
map from the
VERITAS Sky Survey with the associated maps made by the Fermi-LAT
(GeV) and Milagro (multi-TeV) instruments.

\vspace*{0.5cm}
\footnotesize{{\bf Acknowledgment:}{
This research is supported
by grants from the U.S. Department of Energy Office of Science,
the U.S. National Science Foundation and the Smithsonian Institution, 
by NSERC in Canada, by Science Foundation Ireland
(SFI 10/RFP/AST2748) and by STFC in the U.K. We acknowledge 
the excellent work of the technical support staff at the Fred
Lawrence Whipple Observatory and at the collaborating 
institutions in the construction and operation of the 
instrument.
}

\end{document}